%% file: AI_comp_draft.tex
\documentclass[sigconf]{acmart}

\AtBeginDocument{%
  \providecommand\BibTeX{{%
    \normalfont B\kern-0.5em{\scshape i\kern-0.25em b}\kern-0.8em\TeX}}}

\copyrightyear{2023} 
\acmYear{2023} 
\setcopyright{rightsretained} 
\acmConference[AIES '23]{AAAI/ACM Conference on AI, Ethics, and Society}{August 8--10, 2023}{Montréal, QC, Canada}
\acmBooktitle{AAAI/ACM Conference on AI, Ethics, and Society (AIES '23), August 8--10, 2023, Montréal, QC, Canada}
\acmDOI{10.1145/3600211.3604702}
\acmISBN{979-8-4007-0231-0/23/08}

\usepackage[printonlyused,nolist,nohyperlinks]{acronym}
\usepackage{color,soul}
\usepackage[title]{appendix}
\begin{document}
\acrodef{ML}{machine learning}
\acrodef{AI}{Artificial Intelligence}

\title{What does it mean to be a responsible AI practitioner: An ontology of roles and skills}

\author{Shalaleh Rismani}

\affiliation{%
  \institution{McGill University}
  \country{Canada}
}

\author{AJung Moon}
\affiliation{%
   \institution{McGill University}
  \country{Canada}
}

\renewcommand{\shortauthors}{Rismani et al.}

\begin{abstract}
With the growing need to regulate AI systems across a wide variety of application domains, a new set of occupations has emerged in the industry. The so-called responsible \ac{AI} practitioners or \ac{AI} ethicists are generally tasked with interpreting and operationalizing best practices for ethical and safe design of \ac{AI} systems. 
Due to the nascent nature of these roles, however, it is unclear to future employers and aspiring AI ethicists what specific function these roles serve and what skills are necessary to serve the functions. Without clarity on these, we cannot train future \ac{AI} ethicists with meaningful learning objectives.

In this work, we examine what responsible \ac{AI} practitioners do in the industry and what skills they employ on the job. We propose an ontology of existing roles alongside skills and competencies that serve each role. We created this ontology by examining the job postings for such roles over a two-year period (2020-2022) and conducting expert interviews with fourteen individuals who currently hold such a role in the industry. Our ontology contributes to business leaders looking to build responsible \ac{AI} teams and provides educators with a set of competencies that an AI ethics curriculum can prioritize. 
\end{abstract}


\begin{teaserfigure}
  \includegraphics[width=\textwidth]{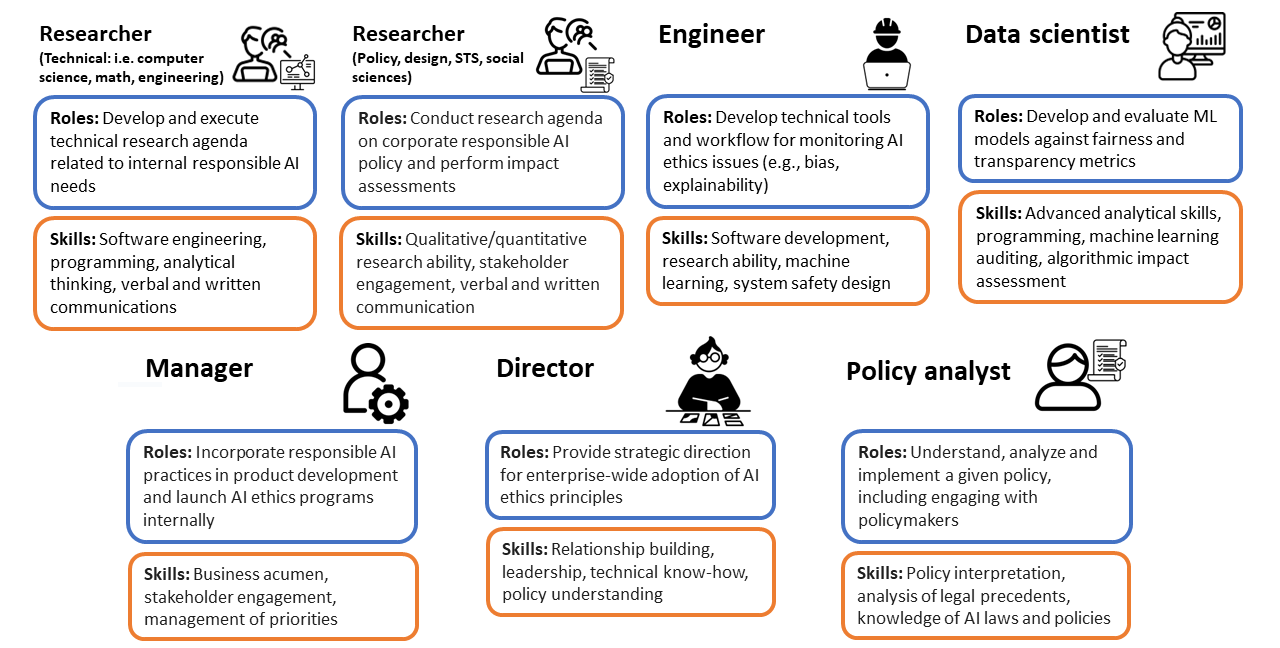}
  \caption{Existing roles and skills expected of responsible AI practitioners (AI ethicists)}
    \label{fig:teaser}
\end{teaserfigure}


\maketitle

\section{Introduction}
With the rapid growth of the \ac{AI} industry, the need for AI and AI ethics expertise has also grown. Companies and governmental organizations are paying more attention to the impact \ac{AI} can have on our society and how \ac{AI} systems should be designed and deployed responsibly ~\cite{Jobin2019,Fjeld2020-rb,Pak-Hang_Wong2020-jv}. From 2015 onward, a series of \ac{AI} ethics principles ~\cite{Jobin2019}, in-depth auditing toolkits ~\cite{Raji2020a, Moon2019, Andersona}, checklists ~\cite{canada, Madaio2020}, codebases ~\cite{microsoft,ibm}, standards and regulations ~\cite{eu, ieee} have been proposed by many different international actors. Several communities of research and practice such as FATE (Fairness, Accountability, Transparency, and Ethics), responsible \ac{AI}, \ac{AI} ethics, \ac{AI} safety and \ac{AI} alignment have emerged. This general movement towards responsible development of \ac{AI} has created new roles in the industry referred to as \textit{responsible AI practitioners} in this paper. The primary mandate of these roles is understanding, analyzing, and addressing ethical and social implications of \ac{AI} systems within the business context. The emergence of these roles challenges technology companies to curate these roles and teams. Leaders in \ac{AI}-related organizations need to identify, recruit and train appropriate candidates for such roles. As the demand to fill such roles continue to increase, educators need effective means to train talent with the right set of skills. 

Recently, scholars examined the common roles responsible AI practitioners serve ~\cite{Wang2023-js,Gambelin2021}, explored the challenges that they face ~\cite{Moss2020, Rakova2021c}, and criticized the problematic nature of the accountability mechanisms that relate to these roles ~\cite{Costanza-Chock2022-ch}. Moreover, others highlight the myriad practical challenges facing the development of a comprehensive training program to fill such roles ~\cite{Raji2021,Borenstein2021,Garrett2020-dw}. However, there is a lack of empirical research investigating the types of roles, corresponding responsibilities, and qualifications that responsible AI practitioners have in the industry. To address these gaps, we examine the following research questions: 

\begin{itemize}
    \item \textbf{RQ1:} What are the types of roles and responsibilities that responsible AI practitioners hold in the industry? 
    \item \textbf{RQ2:} What are the skills, qualifications, and interpersonal qualities necessary for holding such roles? 
\end{itemize}

We address these questions by conducting a two-part qualitative study. We examined 79 job postings from March 2020 to March 2022 and conducted expert interviews with 14 practitioners who currently hold these roles in the industry. Learning from fields of competency-based recruitment and curriculum development, we propose an ontology of different occupations and an accompanying list of competencies for those occupations. 

As illustrated in Figure \ref{fig:teaser}, our competency framework outlines seven occupations that responsible AI practitioners hold in the industry: researcher (of two kinds), data scientist, engineer, director/executive, manager, and policy analyst. For each occupation, the ontology includes a list of responsibilities, skills, knowledge, attitudes, and qualifications. We find that while the roles and responsibilities held by responsible AI practitioners are wide-ranging, they all have interdisciplinary backgrounds and are individuals who thrive in working with individuals from different disciplines. We discuss how educators and employers can use this competency framework to develop new curricula/programs and adequately recruit for the rapidly changing field of responsible AI development.   

\section{Background}

 With the increased media reporting and regulation requirements around social and ethical issues of AI-based products and services ~\cite{Stuurman2022-kb,Mokander2022-ae,aida-euai,Shelby2022-oi,Rismani2023-im,Weidinger2022-ni}, the role of a responsible AI practitioner has emerged as a demanding position in the technology industry. In this section, we provide an overview of debates about these roles and existing educational programs that aim to train future responsible AI practitioners. We discuss how existing competency frameworks treat the role of a responsible AI practitioner and highlight the gaps we address in this work. 

\subsection{Emergence of the responsible AI practitioners}
 Considering the nascency of AI ethics as a domain, only a few scholars have characterized occupations held by responsible AI practitioners ~\cite{Widder2023-lh,Mantymaki2022-im}. For instance, Gambelin frames the role of an \ac{AI} ethicist as "an individual with a robust knowledge of ethics" who has the responsibility and the ability to "apply such abstract concepts (i.e. ethical theories) to concrete situations" for the \ac{AI} system. According to Gambelin, an \ac{AI} ethicist in the industry also needs to be aware of existing policy work, have experience in business management, and possess excellent communication skills ~\cite{Gambelin2021}. Gambelin identifies bravery as the most important characteristic of an \ac{AI} ethicist as they often need to "shoulder responsibility" for potential negative impacts of \ac{AI} in the absence of regulation. 
 
 Moss and Metcalf investigated practices and challenges of responsible AI practitioners in Silicon Valley and described them as "ethics owners" who are responsible for "handling challenging ethical dilemmas with tools of tech management and translating public pressure into new corporate practices" ~\cite{Moss2020}. Echoing Moss and Metcalf's seminal work on examining AI industry practices, a growing body of empirical work highlights that responsible AI practitioners face challenges such as misalignment of incentives, nascent organizational cultures, shortage of internal skills and capability, and the complexity of AI ethics issues when trying to do their day-to-day tasks ~\cite{Schiff2021-et, Nabavi2023-ce, Wang2023-js, Rakova2021c, Rismani2023-im}. Furthermore, only large technology companies often have the necessary resources to hire responsible AI practitioners ~\cite{Sloane2022-ag}. Small and medium-sized companies struggle to access such expertise and rely on openly available information or hire external consultants/auditors as needed ~\cite{Sloane2022-ag,Costanza-Chock2022-ch}. This has given rise to AI ethics as consulting and auditing service ~\cite{orcaa,ethical-advisory,Lab2019-ur}. 
 
 While challenges in operationalizing responsible \ac{AI} practices are an active area of research, there is a gap in understanding the role and necessary competencies of responsible \ac{AI} practitioners in the industry.


\subsection{Qualifications to be a responsible AI practitioner}
The emergence of auditors in the field of responsible AI emphasizes the need for formal training and certification of such roles in the industry ~\cite{Costanza-Chock2022-ch}. This raises a few practical questions: Who is qualified to take these roles? How should these individuals be trained? Are existing computer science, engineering, and social science curricula prepare individuals for such roles? 

Educators responded to this need by developing a range of educational programs and curricula ~\cite{Gorur2020-vu,Williams2021-sd,Quinn2021-jj,Furey2019-xz,Borenstein2021}. 
In a survey of the curricula for university courses focused on AI ethics, Garrett et al. emphasize that such topics should be formally integrated into the learning objectives of current and new courses ~\cite{Garrett2020-dw}. On the other hand, as Peterson et al. describe, discussing social and ethical issues in computer science courses remains a challenge ~\cite{Peterson2023-xa}. They propose pedagogues for fostering the emotional engagement of students in the classroom as a solution ~\cite{Peterson2023-xa}. 

Recognizing the importance of interdisciplinary approaches in AI ethics, Raji et al. argue that computer science is currently valued significantly over liberal arts even in the research area of fairness of machine learning systems ~\cite{Raji2021}. Furthermore, they state that the perceived superiority culture in computer science and engineering has created a "new figure of a socio-technical expert", titled "Ethics Unicorns" - full stack developers, who can solve challenging problems of integrating technology in society.  

This overemphasis on computer science expertise and the trend toward integrating ethics content in existing technical curricula may be problematic if these efforts do not match the skills and disciplinary needs of the industry. It raises questions about whether the educational backgrounds of responsible AI practitioners today are indeed in computer science. In this work, we inform the curriculum development efforts across a diverse range of disciplinary areas by understanding these roles in the industry and outlining the attributes, qualifications, and skills necessary for holding them. 
 
\subsection{Competency frameworks in AI and AI ethics}
\label{competencyframework}
Competency frameworks are useful tools for human resource management (i.e. recruitment, performance improvement) and educational development (i.e. new training programs and curriculum development in universities) ~\cite{CIPD,Spencer1993}. Competency frameworks highlight different competencies required for a profession and link these competencies to skills and knowledge. According to Diana Kramer "competencies are skills, knowledge and behaviours that individuals need to possess to be successful today and in the future" ~\cite{Sanghi2016}. This definition frames our discussion of competency in this paper. 

Competency frameworks help governmental and non-governmental organizations keep track of the type of skills their employees/general public need in the short and long term. Educators use these frameworks to update existing curricula and develop appropriate learning objectives. On the other hand, business leaders and human resource professionals use these frameworks for their recruitment practices.

Today's existing competency frameworks do not sufficiently represent roles and competencies of a responsible AI practitioner. For example, O*NET is United State's national program for collecting and distributing information about occupations ~\cite{Administration}. O*NET-SOC is a taxonomy that defines 923 occupations and they are linked to a list of competencies. Searching the taxonomy for "ethics", "machine learning", "data", "security", and "privacy" leads to minimal results such as "information security analysis", "data scientist and "database architect". The dataset do not include occupation titles such as machine learning engineer/researcher or data/\ac{AI} ethics manager. 

ESCO, the European skills, competencies, qualifications, and occupation is the European and multilingual equivalent of US's O*NET ~\cite{EuropeanCommission2022}. ESCO contains 3008 occupations and  13890 skills. Searching for the above terms leads to more relevant results such as computer vision engineer, ICT intelligent system designer, policy manager, corporate social responsibility manager, ethics hacker, data protection officer, chief data officer, and ICT security manager. However, emerging occupations relevant to \ac{AI} and \ac{AI} ethics have not been well-represented in these established, Western competency frameworks. 

As a response, a number of new \ac{AI} competency frameworks have recently been developed.
One such enabler is the series of projects funded by the Pôle montréalais d’enseignement supérieur en intelligence artificielle (PIA), a multi-institutional initiative in Montreal, Canada aimed to align educational programs with the needs of the \ac{AI} industry. Six projects related to \ac{AI} competency frameworks were funded -- including the work presented in this paper. This resulted in an overarching \ac{AI} competency for postsecondary education that includes ethical competencies ~\cite{Blok2021}, and a competency framework specific to \ac{AI} ethics skills training ~\cite{Bruneault2022}. Bruneault et al., in particular, created a list of \ac{AI} ethics competencies based on interviews of university instructors/professors already teaching courses related to \ac{AI} ethics across North America. 

Our work complements these collective efforts by providing a framework that represents the needs of the industry expressed in recent \ac{AI} ethics-related job postings and the realities of the jobs \ac{AI} ethics practitioners hold in nonprofit and for-profit corporations today.

\section{Methodology}
Practitioners and scholars of different domains typically create competencies frameworks using a process most appropriate for their needs. However, many follow a version of the process highlighted by Sanghi ~\cite{Sanghi2016}. The steps of the process are: 1) Define the purpose and performance objective of a position, 2) Identify the competencies and behaviors that predict and describe superior performance in the job, 3) Validate selected competencies, 4) Implement/integrate competencies and 5) Update competencies. 

In this work, we focus on answering questions raised in the first two steps about the objectives of \textit{responsible AI practitioner} roles and skills/qualities required to perform well in these positions. We take a two-pronged approach to understand the nature of emerging roles under the broad category of \textit{responsible AI practitioners} in the industry. Firstly, we reviewed and analyzed job postings related to our working definition of \textit{responsible AI practitioner}. Secondly, we interviewed individuals who are responsible AI practitioners in the industry today. We then synthesized data collected from these two sources through thematic analyses. We present our proposed competency framework in Section \ref{finding}. This study was approved by the Research Ethics Board of our academic institution.  

\subsection{AI Ethicist Job Postings Review}
We collected and analyzed 94 publicly available job postings over the period of March 2020 to March 2022. The job postings included a range of job titles, including researcher, manager, and analyst.  The following sections describe the process for collecting, selecting, and analyzing these job postings that led to the development of the ontology of responsible AI practitioner roles and skills. 
\subsubsection{Collection of job postings}
 To collect "\ac{AI} ethicist" job postings, we searched and scraped three job-finding websites, including LinkedIn, indeed.com, and SimplyHired, every two months from March 2020 to March 2022. We used the following search terms: \ac{AI} ethics lead, Responsible \ac{AI} lead, \ac{AI} ethics researcher, data OR \ac{AI} ethicist and fairness OR transparency researcher/engineer. Considering that search results only showed a few relevant job postings, we also collected job postings that came through referrals, including mailing lists such as FATML, 80000hours.org, and roboticsworldwide. 
 
 After scanning all the resulting job postings with the inclusion criteria, we gathered a total of 79 job postings for thematic analysis. We included the job postings that were published within our data collection period, were situated in the industry (including not-for-profit organizations), and outlined responsibilities with regards to implementing \ac{AI} ethics practices in a given sector.\footnote{The table outlining the inclusion and exclusion criteria is in the supplemental material.}

\subsubsection{Analysis} 

Using Braun and Clarke's thematic analysis methodology ~\cite{Braun2006-rj}, we analyzed the job postings with the coding scheme illustrated in Table \ref{tab:JP_coding}.  The lead author created this coding scheme after reviewing all the postings. The coding scheme was also informed by frequently used categories across competency frameworks explained earlier in section \ref{competencyframework}. 

The codes were generally split into four key elements: the company environment, responsibilities in the given occupation, qualifications, and skills. The codes of "company environment" and "qualifications - interdisciplinarity" are unique to this coding scheme due to their prevalence in the postings' content. 

After developing the first draft of the coding scheme, a student researcher was trained to use this scheme and coded 10\% of the job postings. The student researcher's analysis using the coding scheme was consistent with the lead researcher's analysis of the same set of job postings. The discussion between the lead and student researcher helped clarify the description and examples for each code. However, there were no new codes that were added to the scheme. The lead author updated the coding scheme and coded the entire set of postings using the new scheme. 

\begin{table}[ht]
\caption{Coding scheme for Job Posting Analysis}
\label{tab:JP_coding}
\begin{tabular}{l}
\textbf{Code}                         \\ \hline
\textbf{Company environment}          \\ \hline
\textbf{Occupation}    \\               \hline
occupation - non-technical roles                      \\ \hline
occupation - technical roles \\ \hline
occupation - title                                     \\ \hline
\textbf{Qualifications}                                      \\ \hline
qualifications - education                                \\ \hline
qualifications - experience  \\ \hline
qualifications - interdisciplinarity  \\ \hline
\textbf{Skills/competency}                             \\ \hline
skills/competency - attitudes/values                   \\ \hline
skills/competency - knowledge  \\ \hline
skills/competency - language skills  \\ \hline
skills/competency - skills \\ \hline
\end{tabular}
\end{table}

\subsection{Expert interviews}
The job postings provide a high-level analysis of the required skills and competencies expressed by recruiters; however, they may not represent the reality of these roles. Therefore, we conducted 14 interviews with experts who currently hold responsible AI practitioner positions in the industry. The focus of the interviews was on understanding the responsibilities, qualifications, and skills necessary for these roles. Considering the objective of this research project on the type of roles and skills, we did not acquire any demographic information about the participants in these roles. This also ensured that we can maintain the anonymity of these participants considering that a limited number of people hold these positions.  

\subsubsection{Recruitment}
We compiled a list of potential interview candidates through (a) referrals within the authors' professional network and (b) we used similar search terms as the ones highlighted for job postings to look for people who currently hold these positions. Moreover, we also considered people from the industry who had accepted papers at relevant conferences such as FAccT and AIES in 2020 and 2021. The suitable participants:
\begin{itemize}
    \item worked for a minimum of three months in their role; 
    \item held this position in the industry or worked mainly with industrial partners;
    \item held managerial, researcher, technical positions that are focused on implementing responsible \ac{AI} practices within the industry. 
\end{itemize}

We did not interview researchers or professors in academic institutes and only interviewed those holding positions at nonprofit and for-profit companies. While we only used the search terms in English to find interview participants for practical reasons, we did not limit our recruitment efforts to a geographical region given the limited number of individuals holding these roles across the industry. We recruited and conducted interviews from June 2021 - February 2022. 

\subsubsection{Interview protocol}
The primary researcher conducted all fourteen interviews. All of the interviews were 45 to 60 minutes in length. The interviewer first described the project and obtained the participant's consent. The interview was semi-structured with ten questions focused on exploring the following four topics:\footnote{The detailed interview protocol is included in the supplementary material.} 
\begin{itemize}
    \item Background and current role
    \item Situation your work, projects in \ac{AI} ethics
    \item Skills, knowledge, values
    \item Looking into the future 
\end{itemize}

\subsubsection{Data Analysis}

The interviews were audio recorded and transcribed. The primary researcher checked and corrected the transcriptions manually afterwards. 
The authors analyzed the interviews deductively and inductively. The lead author applied the coding scheme derived from the job postings (\autoref{tab:JP_coding}) to the interview transcripts. Furthermore, a reflexive thematic analysis ~\cite{Braun2006-rj} was done by the lead author to capture the rich and nuanced details that were not represented in the coding scheme for the job postings. The results from the reflexive analysis were used to contextualize and improve the coding scheme from the job postings. However, no new codes were added to the scheme considering the focus of the research objective addressed in this paper. The lead author presented the coding scheme to the full research team, including a student researcher who had independently reviewed the data to identify general trends. The coding scheme was iteratively revised and refined based on the feedback from the research team. 

\subsection{Author reflexivity and limitations}

We recognize that this research reflects our positionality and biases as academics in North America. Furthermore, the data we collected were all in English and they were representative of job postings and positions in companies situated in North America and Europe. We were not able to collect data on job postings and candidates representing existing efforts in Asia and the Global South. 
Furthermore, we recognize that the roles in this field are continually shifting. Therefore, this ontology is only a snapshot of the roles and skills that responsible AI practitioners have and are recruited for today. Further iterations on these types of frameworks will be necessary in the future as these roles evolve. Finally, this study focuses on examining responsibilities, qualifications, and skills required of today's practitioners independent of their demographic factors (e.g., gender, age). We recognize the importance of representing a demographically diverse group of individuals and their experiences in qualitative research such as ours. Once responsible AI practitioners become a common occupation held by many, future studies should include demographic factors as part of similar investigations.

\section{Proposed competency framework for responsible AI practitioners}
\label{finding}
\begin{figure}[ht]
  \centering
  \includegraphics[width=0.9\linewidth]{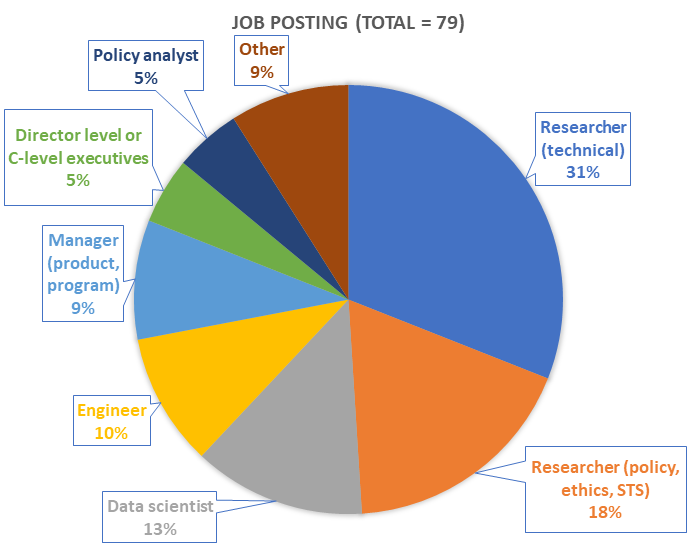}
  \caption{Distribution of occupations represented in the job postings dataset}
  \label{postingchart}
\end{figure}

\begin{figure}[ht]
  \centering
  \includegraphics[width=0.9\linewidth]{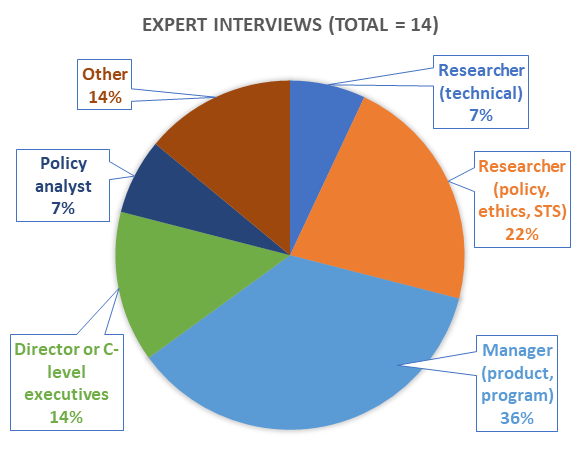}
  \caption{Distribution of occupations represented in the interviews}
  \label{interviewchart}
\end{figure}

From our analysis, we developed a preliminary competency framework that captures seven classes of existing occupational roles and several emerging classes of occupations. Figures \ref{postingchart} and \ref{interviewchart} show how each occupation type was represented in the job postings and interviews. Three of the occupations require technical expertise (researcher, data scientist, and engineer), two require policy expertise (researcher, policy analyst), and the remaining two are managerial (manager, director). In the following sections, we provide a detailed description of the responsibilities, skills, qualifications, and qualities for each of these roles.


\subsection{Researcher (technical)}
The most common class of occupations found in the job postings was that of a researcher focused on technical aspects of fairness, explainability, safety, alignment, privacy and auditability of \ac{AI} systems (24 job postings, 2 interviews). 
Employers represented in this dataset were looking to hire researchers at varying levels of seniority (assistant, associate and principal). 
The main responsibilities of these researchers are split into four main categories: conducting research, communicating their findings, working with other teams (internally and externally), and developing novel solutions for identified problems. 
As expected, research directions set by these researchers need to support company-specific needs, and there is an emphasis on communication between researchers and product, legal and executive teams.  

\paragraph{Skills}
The researchers in this group need to have a mix of technical skills (i.e. software engineering and programming languages such as Python), research skills (i.e. analytical thinking and synthesis of complex ideas ), and leadership skills (i.e. leading and guiding fellow researchers). The dataset from the job postings emphasized equally all these skills, and more senior positions emphasized leadership skills. A senior researcher explained that they look for "different research skills" depending on the project; however, they generally look for \textit{"some background in machine learning, statistics, computer science or something of that nature"} and hire candidates that have some \textit{"interdisciplinary background"}. The data from the postings and the interviews show a strong emphasis on good verbal and written communication skills. Participants highlighted the ability to publish in academic venues and some emphasized the ability to communicate with different audiences internally (i.e. product teams and executives) and externally (policy-makers and executives). A technical researcher emphasized the importance of\textit{ "convincing stakeholders"} and creating \textit{"strategic collaborations"} by communicating with practitioners with "diverse" backgrounds. 

\paragraph{Qualifications} 
The job postings mainly aim to attract candidates who have a PhD in computer science or a related field. Few of the job postings accept a master's in these fields, whereas some do not highlight a specific degree and mainly focus on necessary skills and knowledge. The majority of postings have a heavy emphasis on the required experience. Interview participants also emphasized the importance of experience. A research manager expressed that they are not necessarily looking for a \textit{"PhD in computer science"}. They are looking for candidates with experience in \textit{"leading and executing a research agenda"}, working with different people and teams, synthesizing and \textit{"communicating challenging concepts"}, and practicing software engineering. Some postings highlight experience with implementing \ac{AI} ethics-related concepts. However, this was often listed as a preferred qualification rather than a required one. Similarly, researchers we interviewed, echoed the importance and value of having a publication record in \textit{"Fairness, Accountability, Transparency, and Ethics (FATE) communities"} such as ACM Conference on Fairness, Accountability, and Transparency (FAccT) and AAAI/ACM Conference on AI, Ethics, and Society (AIES). 

\paragraph{Interpersonal Qualities}
The most common attitude/value was the aptitude and interest to \textit{collaborate and work in an interdisciplinary environment}. A researcher emphasized that the current conversations are "engineering focused" and they actively incorporate perspectives from\textit{ social science and philosophy by collaborating with experts in these areas}. The most desired value was \textit{"curiosity to learn about [responsible AI] problems"}. Many of the participants highlight other values and attitudes such as \textit{"passion"} towards building safe and ethical \ac{AI} systems, \textit{willingness to manage uncertainty and challenges}, \textit{creativity}, and \textit{resourcefulness}.

\subsection{Data scientist}
The data scientist occupation is represented in 10 job postings in our dataset, and none in the interviews. 
The job postings seek to fill traditional data scientist roles with an added focus on examining responsible \ac{AI}-related issues. The common responsibilities outlined for these positions are a) to collect and pre-process data, and b) to develop, analyze, and test models -- these are typical of existing data science roles. However, the job postings emphasize the position's responsibility to test machine learning models for \ac{AI} ethics concerns such as fairness and transparency. Data scientists who work in the responsible AI domain have additional non-conventional roles. These roles include understanding and interpreting existing regulations, policies, and standards on the impact of \ac{AI} systems and testing the systems' capability for elements covered in these policies. They also need to work with technical and non-technical stakeholders to communicate findings, build capacity around responsible AI concepts and engage them as needed. 

\paragraph{Skills}
The job postings put a heavy emphasis on advanced analytical skills and the ability to use programming languages such as R, Python and SQL for basic data mining. The ability to learn independently in a new domain and master complex code base is also listed as one of the key skills. A few of the postings list project management and organizational skills; however, this is not common. When it comes to the knowledge required, the focus shifts from the technical domain to an understanding of fields such as sociology, critical data studies, and AI regulations. Many postings highlight that potential candidates need to be familiar with concepts such as \ac{AI}/ML auditing, algorithmic impact assessments, assessment of fairness in predictive models, explainability, robustness, and human-\ac{AI} interaction. Technical knowledge, such as understanding transfer-based language models and logistic regression model development, is also highlighted in the posting. Lastly, the job postings outline the need for strong interpersonal, verbal, and written communication skills. However, experience publishing and presenting at academic venues is not mentioned. 
\paragraph{Qualifications} 
The majority of the job postings require a bachelor's degree in quantitative fields such as data science and computer science and prefer higher degrees (master's or Ph.D.). Companies are looking for candidates who have experience in data science, software engineering, and worked with large language models. Moreover, they are looking for experience in putting responsible \ac{AI} principles into practice, evaluating the ethics of algorithms, and having basic familiarity with law and policy research. The ability and experience to translate \ac{AI} ethic principles into practice are heavily emphasized throughout these job postings. 

\paragraph{Interpersonal Qualities}
 The job postings emphasize the ability to work with people from different backgrounds. However, these job postings do not include a comprehensive list of values. A few postings mention being a self-starter, working collaboratively to resolve conflict, and caring deeply about the data used to train ML models as key attitudes. Being flexible, innovative, curious, adaptive, and passionate about tackling real word challenges are also some of the sought-after values.

\subsection{Engineer}

The engineer occupation is represented in 8 of the job postings. None of our interview participants belong to this category. The key responsibility of an engineer practicing \ac{AI} ethics is to help establish a safety culture and system within an organization by developing technical tools. They are tasked with developing a workflow for modeling and testing for issues such as bias, explainiability, safety, and alignment of \ac{AI} systems. As part of this, engineers need to create code bases that could be used across the \ac{AI} system development pipeline based on existing and evolving best practices.

\paragraph{Skills and Qualifications} 
Job postings for engineers place a significant emphasis on experience-based qualifications and skills. The companies represented in this dataset are looking for skills and experience in software development, dataset production pipelines, researching fairness and safety implications of ML systems, and the development of large language models. They are also looking for experience working in a fast-paced technology company. Based on these qualifications, the main set of skills are programming and \ac{AI}/ML development skills and this needs to be supported by knowledge and familiarity with foundational concepts in \ac{AI}/ML, fairness, explainability, system's safety, and safety life cycle management. Lastly, most of the job descriptions do not have a heavy emphasis on communication skills. Only a few mention excellent written and oral communication skills as a requirement.

\paragraph{Interpersonal Qualities}
 In contrast to the lack of emphasis on communication skills, these postings have a particular focus on the attitude and values of ideal candidates more so than any other occupation category. These attitudes include being result-oriented, willingness to contribute as needed (even if not specified) and keen to learn new concepts. They are looking for people who value working on challenging problems and care about the societal impact of their work. 

\subsection{Researcher (law, philosophy and social sciences)}
The second most frequent category of postings belongs to researchers that focus on topics such as policy, sociotechnical issues, and governance (14 job postings, 3 interviews). We created a separate group of positions as their responsibilities, skills, and qualities are sufficiently different from the technical researcher position. Candidates in this category need to conduct research, perform ethics or impact assessments of \ac{AI} systems, act as a liaison and translator between research, product, policy, and legal teams, and lastly, advise on policy, standards, and regulations-related matters internally and externally. When conducting research, two different focus areas come up in the job postings: testing and evaluating \ac{AI} system to inform policy and researching existing policies/regulations, and translating them into practice.
\paragraph{Skills}
The job postings highlight two sets of distinct skills for this group of researchers. Firstly, these researchers require a basic level of programming, advanced analytics, and data visualization skills. Few positions highlighted the need for even more advanced ML and \ac{AI} skills. It is noteworthy that despite these researchers' focus on policy, governance and sociotechnical issues, the postings still require them to have some data analytic skills. Secondly, these researchers need to have excellent facilitation, community-building, and stakeholder engagement skills. These two skills need to be complemented by strong leadership and management skills.
The job postings heavily emphasize strong communication skills for this group of researchers. Besides the conventional skill of presenting and publishing papers, this group of researchers need to effectively work across different functionalities and disciplines.
On a similar trend, these researchers need to have expertise in a variety of areas. They need to have a good understanding of \textit{"qualitative and quantitative research methods"}, reliably know the current and emerging\textit{"legal and regulatory frameworks and policies"}, be \textit{"familiar with \ac{AI} technology"} and have a good knowledge of practices, process, design, and development of \ac{AI} technology. This is a vast range of expertise and often \textit{"very difficult to recruit"} for as highlighted by our expert interviewees. 

\paragraph{Qualifications} 
Just over half of the job postings list a Ph.D. in relevant areas as a requirement, including human-computer interaction, cognitive psychology, experimental psychology, digital anthropology, law, policy, and quantitative social sciences. Two postings require only a bachelor's or a master's in the listed areas. Similar to the technical researcher occupation, some positions do not specify any educational requirements and only focus on experience and skills. Our expert interviewees in this category are from a range of educational backgrounds ranging from a master's in sociotechnical systems, a law degree combined with a background in statistics, and a master's in cognitive systems. 

Besides experience in research, companies are looking for experience in translating research into design, technology development, and policy. A researcher explained that they need to do a lot of "translational work" between the academic conversation and product teams in companies. A good candidate for this occupation would have \textit{"project management"}, \textit{"change management"}, \textit{"stakeholder engagement"}, and \textit{"applied ethics"} experience in a "fast-paced environment". All four of these skills do not appear in all of the job postings and interview discussions. However, a permutation of them appears throughout the job posting data and participants' responses. 

\paragraph{Interpersonal Qualities}

As emphasized strongly in both of the datasets, ideal candidates in this category need to have a \textit{"figure-it-out somehow"} or \textit{"make it happen"} attitude as explained by a participant. They are \textit{"driven by curiosity and passion towards"} issues related to responsible AI development and are excited to engage with the product teams. Participants noted that ideal candidates in these roles are \textit{"creative problem solvers"} who can work in a \textit{"fast-changing environment"}.

\subsection{Policy analyst}
Policy analyst occupation is the least represented [1 expert interview, 4 job postings] in our data sources; however, considering the consistent list of competencies, we decided to include it within the proposed framework. The role of a policy analyst is to understand, analyze and implement a given policy within an organization. Moreover, they need to engage with policymakers and regulators and provide feedback on existing policies. 

\paragraph{Skills and Qualifications} 
A policy analyst needs to have proven knowledge of laws, policies, regulations, and precedents applicable to a given technology when it comes to \ac{AI} ethics-related issues. Moreover, all of the job postings highlight the importance of familiarity with \ac{AI} technology. According to the job postings, a good candidate would have experience in interpreting policy and developing assessments for a given application. They also need to be skilled in management, team building, and mentorship. This finding echoes remarks from expert interviews. Even though none of the job postings specify an educational degree requirement, the expert we interviewed was a lawyer with a master's in technology law. 

\paragraph{Interpersonal Qualities}
The job postings in this category heavily emphasized values and attitudes. A good analyst needs to have sound judgment and outstanding personal integrity. They should be caring and knowledgeable about the impact of technology on society. Moreover, they enjoy working on complex multifaceted problems and are passionate about improving governance of \ac{AI} systems. The expert interviewee's perspective closely matches these attributes. Participants elaborated that they needed to be \textit{"brave"} and \textit{"step up to ask questions and challenge status quo consistently over a long time"}. As expected communications skills are considered critical for success. The expert interviewee significantly emphasized the importance of \textit{"networking as a key factor"} in succeeding in their role. 

\subsection{Manager} 
We analyzed 7 management-related job postings and 5 expert interviewees in this category. The product managers take the role of incorporating responsible \ac{AI} practices in the product development process. In contrast, program managers are often leading and launching a new program on establishing \ac{AI} ethics practices within the organization. These programs often involve building an organization's capacity to manage responsible \ac{AI} issues. 

\paragraph{Skills} 
For both streams of management, the potential candidates need to have strong business acumen and a vision for the use/development of AI technology within an organization. Some of the key management skills highlighted in the job postings include the ability to manage multiple priorities and strategically remove potential blockers to success. Another sought-after skill is the ability to effectively engage stakeholders in the process. Expert interviewees also echoed the importance of this skill as their roles often involve getting people \textit{"on board with new ways of thinking and creating"}. According to the job postings, good candidates for management need to have a practical understanding of the \ac{AI} life cycle and be familiar with integrating responsible \ac{AI} practices into a program or a product. Our interviewees note that they continuously need to \textit{"learn and keep up with the fast-paced development of \ac{AI}"}.

\paragraph{Qualifications} 
Not many postings have highlighted educational qualifications and instead focused on experience qualifications. However, the main educational qualification is a bachelor's degree with a preference for higher degrees. The postings have primarily highlighted a degree in a technical field such as computer science or software engineering. Interestingly the interviews reflect a different flavor of educational backgrounds. All of the experts we interviewed had at minimum a master's degree and the majority of them completed their studies in a non-technical field such as philosophy, media studies, and policy. However, these individuals had acquired a significant level of expertise in \ac{AI} ethics through \textit{"self-studying"} and \textit{"engaging with the literature"} and the responsible AI \textit{"community"}. For example, two of the participants trained in technical fields and had a significant level of industry experience. Similarly, they had learned about responsible \ac{AI} through their own initiative. 

On the other hand, the job postings heavily focus on \textit{experience}, including a significant amount of technical know-how, experience focused on ML development, product and program management , and implementation of ethical and social responsibility practices within fast-paced technology companies. The interview participants had been \textit{"working in the industry for some time"} before taking on these management roles. However, their range of experiences do not cover all of the required experiences outlined in the job descriptions. As expected, excellent communication skills are noted in the job descriptions and strongly echoed by the experts as well. The job postings do not necessarily elaborate on the nature of communication skills; however, the experts note that the \textit{"ability to listen"}, \textit{understand}, and sometimes 
"\textit{persuade different stakeholders"} is key in such roles.

\paragraph{Interpersonal Qualities}
 Few of the job postings make remarks about attitudes/values and highlight that managers need to value designing technology for social good and cooperation with other stakeholders. A good candidate for management should  foster a growth mindset and approach their work with agility, creativity, and passion. All of the participants expressed their passion for developing ethical technology  and indicate that they took a lot of initiative to learn and contribute to the field within their company and externally before they could take on their management roles. 

\subsection{Director}

The job descriptions dataset has 4 postings for director positions and 2 of the expert interviewees have directorship roles. According to the job postings, director responsibilities include at least three of the following: a) lead the operationalization of \ac{AI} ethics principles, b) provide strategic direction and roadmap towards enterprise-wide adoption and application of ethical principles and decision frameworks, and c) build internal capacity for \ac{AI} ethics practice and governance. Depending on the nature of the organization and its need to incorporate \ac{AI} ethics practices, these responsibilities vary in scope. For example, a director within a technology start-up will only be able to commit \textit{"limited amount of time to operationalizing \ac{AI} ethics principles and building internal capacity"} compared to a director within a larger technology company. 

\paragraph{Skills and qualifications} 
According to the job postings, the key skill for being a director is having the ability to build a strong relationship with a broad community that helps define and promote best practice standards of \ac{AI} ethics. An ideal director can effectively pair their technical skills/know-how with their management skills and policy/standards knowledge to develop strategic plans for the company. Experience in directing and leading teams, particularly in social responsibility practices within technology companies is highly valued for such positions. Only one job posting specifies an educational (a bachelor's related to policy development and implementation). Others only highlight experience. The two interviewees hold master's degrees in business and information systems respectively. They also had extensive industry experience that was not directly in \ac{AI} ethics. However, their experience involved \textit{"translation of policy within a technology application"}. 

\paragraph{Interpersonal Qualities}
As expected, according to the job postings a good candidate for directorship needs to have exceptional written and verbal communication skills, need to be able \textit{"to articulate complex ideas"} to technical and non-technical audiences, \textit{"engage and influence stakeholders"} and \textit{"collaborate with people from different disciplines, and cultures"}. This set of skills was reflected in our expert interviews. Both interviewees emphasized how they maintain a good flow of communication with the employees and how they remain always open to having conversations on a needs basis. This allowed them to build trust within the company and pursue moving forward with their strategic plan. The job postings highlight the ability to earn trust in relationships as a sought-after value for a directorship role. A director should also be able to \textit{challenge the status quo, be passionate about good technology development, be comfortable with ambiguity}, and \textit{adapt rapidly to changing environment and demands}. Most importantly, a director needs to have \textit{"a strong and clear commitment to the company values"} as they set the tone for others within the organization. 
\subsection{Emerging occupations}

Besides the abovementioned classes of occupations, we found a few other positions that do not map easily to any of the existing categories. 
Considering the limited number of these positions, they do not justify a category of their own. However, we note these emerging roles to understand how they might shape up the responsible AI profession. These occupation titles include data ethicists (2 in job postings), \ac{AI} ethics consultants (2 in interviews), dataset leads (2 in job postings), communication specialist (1 in job postings), safety specialist (1 in job posting) and UX designer (1 in job postings). The following describes the main function of these positions: 
\begin{itemize}
    \item Data ethicist: manage organizational efforts in operationalizing \ac{AI} ethics practices through policy and technology development work. This role has similarities to the role of a policy analyst and data scientist. 
    \item \ac{AI} ethics consultant: apply their expertise in \ac{AI} ethics to solve pain points for consulting clients.
    \item Dataset lead: curate datasets while accounting for fairness and bias-related issues. 
    \item Safety specialist: use and test large language model-based systems to identify failures and errors. 
    \item \ac{AI} ethics communication specialist: write communication pieces that focus on \ac{AI} ethics issues. 
    \item UX designers: design user interfaces with ethics in mind.
\end{itemize}

\subsection{Future of the responsible AI profession}
Our interview participants shared a variety of responses to the question "what will the future of their job be like?". 
Some participants thought that eventually, \textit{"everyone in a company will be responsible"} for understanding ethical and social issues of \ac{AI} as part of their job. In this scenario, everyone would need to have the appropriate knowledge and skillset to apply responsible \ac{AI} practices in their work or at least know when they need to ask for advice from internal or external experts. 

On the contrary, many participants expressed that \textit{"dedicated roles"} need to be recruited. These participants elaborate that recruitment for these roles is and will \textit{"continue to be challenging"} as it is difficult to find people with interdisciplinary backgrounds and established industry work experience. Many of the managers we interviewed have chosen \textit{"to build teams that come from different disciplinary backgrounds"} and provide \textit{"professional development opportunities"} on the job. However, they also described that hiring people into these roles is challenging since corporate leaders are not always willing to invest a lot of resources in \ac{AI} ethics. This often can lead to \textit{"exhaustion and burn-out"} for individuals who currently hold these roles - this is especially true for small and medium-sized technology companies. According to participants, this will likely change with a progressive shift in the regulatory landscape.

\section{Discussion}
Educators and employers play a pivotal role in shaping a responsible AI culture. In our efforts to create a competency framework that outlines the range of roles for responsible AI practitioners, we find that such frameworks can not only guide corporate leaders to recruit talent but also help grow their responsible AI capacity.

We find that the ability to work in an interdisciplinary environment, communicate and engage with diverse stakeholder groups, and the aptitude for curiosity and self-learning are consistently highlighted for all of the roles.
This emphasizes the need to foster an environment where students and existing employees in different roles are encouraged to adopt interdisciplinary approaches/collaboration and explore responsible AI content.

In this section, we articulate how an interdisciplinary environment can be fostered, the importance of organizational support for responsible AI practitioners, and the need to proactively monitor the rapidly changing occupational demand and landscape for these roles.

\subsection{Being able to work in an interdisciplinary environment is critical}

Our results show that many of the responsible AI practitioners today come from non-traditional, non-linear, and interdisciplinary educational and work backgrounds to their current positions. The educational and work experiences of these participants span a multitude of fields and allowed them to develop a strong set of skills in navigating disciplinary boundaries and understanding problems from diverse perspectives. The participants often described their role as a \textit{translator} and \textit{facilitator} between different groups and disciplines within the organization. For instance, they remarked that a concept such as fairness, transparency, or ethically safe has completely different meanings depending on the personal and professional backgrounds of their audience. The participants often needed to translate what these concepts mean across different disciplinary boundaries (i.e. statistics and law).

Notably, while the job postings asked for a diverse array of skills and qualifications from multiple disciplines, those who hold such positions today are often specialized in one or two disciplines. However, they had been exposed to and worked across multiple disciplines in their professional career. The most important asset that our interviewees emphasized was being able to work across disciplinary boundaries. The candidates who successfully hold such positions are\textit{ not "ethical unicorn, full stack developers"} ~\cite{Raji2021}. However, they have honed the skills necessary to translate and create solutions to responsible AI issues across multiple disciplines. Building on existing proposal to improve responsible AI practices ~\cite{Madaio2020,Rismani2023-im,Costanza-Chock2022-ch} and education ~\cite{Garrett2020-dw,Peterson2023-xa}, we posit that AI team leaders need to pay a special attention to hiring individuals with the capability to create, critique and communicate \textit{across multiple disciplines}. Consider Furthermore, educators can get inspiration from education models in highly interdisciplinary fields such as healthcare and create curricula/spaces where students work with peers from different academic backgrounds ~\cite{Klaassen2018-tk,Dyer2003-bd,Hall2001-rg}.

\subsection{Responsible AI practitioners are advocates - but they need organizational support}

We find that responsible AI practitioners are often highly driven and motivated to make a positive impact. These individuals often hold a strong sense of valuing social justice and want to ensure that \ac{AI} technology is developed in a way that is good for society's well-being. One of the most consistent ideas that came through in the interviews is the attitude that the participants had toward their careers. Many of the interview participants took the time to immerse themselves in learning new topics and expressed that they were self-motivated to do so. This is especially true for the individuals who are taking some of these first positions in the industry. When looking at the career trajectory of many of the participants, we observe that they often created their own roles or came into a newly created role. Moreover, these individuals often needed to start their own projects and create relationships with others in the organization to measure their own progress and establish credibility. 

Similar to any emerging profession many of the participants act as champions for ethical and safe development of \ac{AI}. They are often working in an environment that questions and challenges the need for considering \ac{AI} ethics principles. As some of the participants remarked, they often have to answer questions such as "\textit{why do we need to pay for ethics assessments?"},\textit{ "what is the value of considering \ac{AI} ethics in a start-up?"}, or \textit{"why should we put in the time? what is the value added?"}. This act of advocating for \ac{AI} ethics is even more challenging when existing regulations do not have proper enforcement mechanisms for responsible AI practices ~\cite{Costanza-Chock2022-ch}. Many of the participants assume the role of an advocate and often use their excellent communication skills to build relationships and capacity within their organization. 

For the successful implementation of responsible AI practices, it is important that business leaders pay attention and support the advocacy efforts of these practitioners. Many of today's responsible AI practitioners are working with limited resources ~\cite{Moss2020}, have critical responsibilities ~\cite{Rakova2021c}, and are experiencing burn-out ~\cite{Heikkila2022-ld}. Whenever possible, leaders in AI companies need to create appropriate incentive structures, provide the necessary resources and communicate the value of establishing responsible AI practices to their employees so that these practitioners have the necessary support for the effective execution of their responsibilities. Recognizing the nature of these roles, educators can learn from existing methods ~\cite{Farr2009-ab,Carter2011-np} and integrate leadership training into their curricula when addressing responsible AI-related content. 
\subsection{Educators and employers need to monitor and plan for the rapidly changing landscape of responsible AI roles}

The nature of occupations in the AI industry is continually growing and shifting. The rapid technological development ~\cite{Meek2016-wz,mit_2021-dg}, upcoming regulations ~\cite{aida-euai} and global economic conditions ~\cite{Goldman2022-ip,Knight2022-ym} impact how companies recruit and retain responsible AI expertise. Furthermore, there is a need for new educational efforts and programs for preparing new graduates to take on responsible AI practices. The proposed ontology provides a synthesis of roles that have emerged in responsible AI practice and it can serve as a planning tool for corporate leaders and educators. 

Corporate leaders can use this ontology to build internal capacity for individuals who currently hold researcher, data scientist, engineer, policy advisor, manager, and director roles in their institutions. Depending on these companies' responsible AI needs and resources, business executives can work towards creating interdisciplinary teams for establishing responsible AI practice by recruiting individuals with the competencies outlined for each of these roles. Besides recruiting and fostering for responsible AI competencies, these leaders need to communicate the importance of these practices and start by creating the appropriate organizational incentives and resources for adapting responsible AI practices. Government and non-governmental organizations could support such efforts, particularly small and medium-size companies, by formally recognizing such roles in their taxonomies of occupations ~\cite{Administration,EuropeanCommission2022} and providing resources ~\cite{Bessen2022-gy}. 

Current computer science and engineering education focuses primarily on teaching professional ethics ~\cite{Peterson2023-xa}. There is minimal focus and resources on cultivating skills and knowledge required for cultivating the skills that focus on ethics in design~\cite{Garrett2020-dw}. On the other hand, there is a lack of clarity of how much students in social and political sciences need to work on their technical acumen to become skilled responsible AI practitioners \cite{Raji2021}. Educators could use the list of competencies to develop a set of learning objectives and examine the efficacy of different teaching pedagogies in supporting these objectives. Moreover, Educators can use the competency framework as a tool for acquiring resources for further curricula and program development. 

 Notably, the proposed ontology primarily focuses on type of roles, responsibilities and skills without addressing other important factors in recruitment and education efforts such as diversity of individual who get to learn about responsible AI issues or take such roles in the industry. Therefore, it is critical that users of this ontology, consider factors that are not captured in the scope of this ontology. Furthermore, considering the rapidly changing conversation around responsible AI practices, the type of roles in this onotlogy will shift and expand. We invite the community of researchers , practitioners and educators to reflect on these roles and build on this ontology.

\section{Conclusion}

With the increased regulatory activities in the industry, companies have the incentive to ensure responsible AI development. In this work, we found seven different type of roles and their corresponding responsibilities, skills, qualifications, and interpersonal qualities expected in today's responsible AI practitioner. We propose a preliminary competency framework for responsible AI practitioners and highlight the importance of creating interdisciplinary teams and providing adequate organizational support for individuals in these roles. 

\begin{acks}
We thank our study participants for taking the time to share their experiences, expertise, and feedback. We also thank our anonymous reviewers for their deep engagement and valuable feedback on this paper. This work benefitted greatly from the data collection and analysis assistance from our collaborators Sandi Mak, Ivan Ivanov, Aidan Doudeau, and Nandita Jayd at Vanier College, Montreal. We are grateful for their contributions. Finally, this work was financially supported by the Natural Sciences and Engineering Research Council of Canada and Pôle montréalais d’enseignement supérieur en intelligence artificielle. 
\end{acks}

\bibliographystyle{ACM-Reference-Format}
\bibliography{sample-base}
\begin{appendices}
\input{appendix}

\end{appendices}
\end{document}

%% file: appendix.tex
\newpage
\onecolumn
\section{Supplementary Material}
\label{sec:supplementary}
\subsection{Interview protocol}

\paragraph{Consent Process    
}Thank you for reading and signing the Human Subjects Consent Form for this project..

\paragraph{Introduction}  
Thank you for agreeing to take part in this study. My name is [interviewer] and I will be conducting this interview. I am a research assistant working with [advisor].
We have invited you to take part in this today because of your current role. The purpose of this study is to examine the experiences of professionals including ethicists, technologists and business leaders who are dealing with ethical and social implications of particular AI technology through development and implementation. 

Today I am playing two roles: that of a interviewer and that of a researcher. 
At this time I would like to give a brief overview of the project and the consent form. [5min] 

\textbf{Background and current role 
}
\begin{itemize}
    \item Please tell us about your role at your current company. 
\item What is your official job title, and what are you main responsibilities? 
\item Could you please tell me about your background, expertise and experience that led you to take on your current role? 
\item Who do you work most closely with at your company? Who do you manage?  Who do you report to? \item Who are clients? Who are your partners? 
\end{itemize}

\textbf{ 
Situation your work, projects in AI ethics 
}
\begin{itemize}
    \item How do you situate your work within the broader field of AI ethics? What types of challenges are you working on in this field? Please feel free to share any specific examples from your projects.
\item What are the main projects related to AI ethics that you are working on that you can tell us about?
What types of resources (academic papers, academic experts, standards, guidelines) do you use in your AI ethics practice?
\item Do you use the guidelines on AI ethics practice that have published in this field over the past 5 years? If so, which one and what does following the guideline look like at your company? 
\item What are the most important challenges to implementing ethics principles at your work?

\end{itemize}

\textbf{Skills, knowledge, values 
}
\begin{itemize}
    \item What are the most important skillsets, knowledge base and values that you currently use at your job? What are you currently developing and will need in the future? 

\item If you decide to hire someone to replace you in your current role, what would you look for? What skills or background would your ideal candidate have? 

\end{itemize}

\textbf{Looking into the future
}
\begin{itemize}
    \item From your perspective, what roles do you think are necessary in the field of AI ethics in academia, industry, governmental and non-governmental organizations? Please elaborate. 

\end{itemize}
\newpage
\subsection{Relevant tables}

\begin{table*}[h]
\caption{Coding scheme for Job Posting Analysis}
\label{tab:JP_coding_full}
\begin{tabular}{p{0.3\textwidth}p{0.3\textwidth}p{0.3\textwidth}}
\textbf{Code}                        & \textbf{Description}                                          & \textbf{Example}                                   \\ \hline
\textbf{Company environment}         & Company culture and values                                    & "The X company values integrity and public safety" \\ \hline
\textbf{Occupation}                  & Roles and responsibilities                                    &                                                    \\ \hline
occupation - non-technical roles     & Non-technical roles and responsibilities                      & "Manage an interdisciplinary team"                 \\ \hline
occupation - technical roles &
  Technical roles and responsibilities &
  "Create machine learning models that incorporate f\ac{AI}rness metrics" \\ \hline
occupation - title                   & Title in the job posting                                      & "Sociotechnical researcher"                        \\ \hline
\textbf{Qualifications}              & Items highlighted in the qualification section of job posting &                                                    \\ \hline
qualifications - education           & The educational background and tr\ac{AI}ning                       & "PhD in computer science"                          \\ \hline
qualifications - experience &
  Experience related to previous work, projects and volunteer work &
  "At least five years of management experience" \\ \hline
qualifications - interdisciplinarity &
  Experience or education in interdisciplinary environment/setting &
  "Has the ability to work with people from different backgrounds" \\ \hline
\textbf{Skills/competency}           & What are their skills/competencies?                           &                                                    \\ \hline
skills/competency - attitudes/values & how a candidate would approach their work                     & "Looking for a confident leader"                   \\ \hline
skills/competency - knowledge &
  the candidate's required knowledge and understanding &
  "Have a deep understanding of existing \ac{AI} ethics practices" \\ \hline
skills/competency - language skills &
  the candidate's communications skills &
  "Can communicate in a clear and concise way" \\ \hline
skills/competency - skills &
  the candidate's management, research and technical skills &
  "Has the ability of initiate and manage a full research agenda"
\end{tabular}
\end{table*}

\begin{table*}[h]

\caption{Inclusion and exclusion criteria for job postings}
\label{tab:JP_criteria}

\begin{tabular}{p{0.4\textwidth}p{0.4\textwidth}}
\hline
Inclusion criteria              & Exclusion criteria                                                                                        \\ \hline
Between March 2020 - March 2022 & Job postings before March 2020 were not included as it is difficult to find job postings retroactively. \\ \hline
The role requires the potential candidate to work with industry partners or within a given industry &
  The role only requires the candidate to interface with academia (i.e. professors and post-docs that only work in academia were not recruited) \\ \hline
The role requires the candidate to implement \ac{AI} ethics practices (this could be within management, technical or applied research context) &
  The role only involves research in \ac{AI} ethics and does not focus on implementation. \\ \hline
\end{tabular}
\end{table*}

\begin{table*}[]
\caption{Competency framework for \ac{AI} ethics}
\label{tab:framework}
\resizebox{\textwidth}{!}{%

\begin{tabular}{p{0.25\textwidth}p{0.25\textwidth}p{0.25\textwidth}p{0.25\textwidth}}
\textbf{Occupation}                            & \textbf{Titles}                                                                                                                                                                                                                                    & \textbf{Responsibilities}                                                                                                                      & \textbf{Skills}                                                                                                                                                                                                                                                                                                      \\ \hline
\textbf{Researcher (technical)}           & (Senior)   research associate, research assistant, applied researcher, research   scientist, postdoctoral researcher, research fellow, principle applied   scientist                                                                               & 1) Develop and execute research agenda to contribute to FATE and responsible AI community 2) Conduct research that addresses internal responsible AI challenges  
3) Communicate research findings with teams internally and external research community &
1) Software egineering and programming 2) Research skills such as analyitical thinking and synthesis of complex ideas 3) Leading and guiding fellow researchers 4) Good verbal and written communications                                        \\ \hline
\textbf{Researcher (policy, ethics, STS)} & Research   scientist, research associate, senior researcher, fellow, postdoctoral   researcher                                                                                                                                                     & 1) Develop and execute research agenda to contribute to FATE and responsible AI community 2) Perform ethics or impact assessments on internal products 3) Advice on policy, standards and regulations internally and externally 4) Act as a liason and translate policies in practice &
1)Qualitative and quantitative research skills 2) Facilitation, community and stakeholder engagement 3) Communitcation skills 4) Know the current and emerging legal and regulatory frameworks and policies 5)Be familiar with AI technology and its development process \\ \hline
\textbf{Data scientist}                    & Data   scientist compliance officer, data scientist in X, (senior/principal) data and applied scientist,  data science contractor, data scientist, staff data scientist                                                                        & 1) Collect and pre-process data 2) Develop and evaluate ML models 3) Evaluate models for ethics concerns such as fairness and transparency 4) Understand and interpret existing regulations and policies 5) Communicate findings across different groups 6) Build capacity around responsible AI topics                                             & 1) Advanced analytical skills 2) Programming languages such as R, Python and SQL 3) Learn and master a complex code base 4) Familiar with concepts such as ML auditing, algorithmic impact assessments, etc.                                                                                                            \\ \hline
\textbf{Engineer}                          & Research/ML/senior   engineer                                                                                                                                                                                                                      & 1)Develop technical tools to establish safety system and culture in the organization 2) Develop a workflow for modelling and testing for issues such as bias, explainability, safety and alignment &   - 1) Software development 2) Research skills in responsible AI and FATE fields 3) Knowledge and familiarity with foundational concepts of ML, FATE and system safety                                                                                                                                                                                                                                                        \\ \hline
\textbf{Director or C-level executives}  & Chief   of responsible \ac{AI} program, director of data usability and ethics, director   product management, lead of data governance,   director of data ethics and governance, executive director, chief operating   officer & 1) Provide strategic direction and roadmap towards enterprise-wide adoption of AI ethics principles 2) Build internal capacity for responsible AI practice and governance & 1) Build relationship with the broad community 2) Technical know-how 3) Management skills 4) Knowledge of policy and standards 5) Directing and leading team                                                                                                                                                          \\ \hline
\textbf{Manager (product, portfolio)}    & Senior   principal product manager, director of product, program manager, solutions   lead, senior manager of \ac{AI} risk, head of product, venture manager                                                                                            & 1) Incorporate responsible AI practices in product development 2) Lead and launch a new program on establishing responsible AI practices 3) Build internal capacity to manage responsible AI issues &
1) Strong business acumen 2) Manage priorities and blockers to success 3) Engage stakeholders throughout a process 4) Practical understanding of the AI life cycle 5) Ability to keep updated with fast-paced development of AI\\ \hline
\textbf{Policy analyst}                  & Ethics   policy analyst, head of public policy, senior technical policy analyst                                                                                                                                                                    & 1) Understand, analyze and implement a given policy within an organization 2) Engage with policymakers and regulators 3) Provide feedback on existing policies & 1) Proven knowledge of laws, policies, regulations, and precedents applicable to a given technology when it comes to AI ethics-related issues 2) Experience in interpreting policy and developing assessments for an application 3) Skilled in management, team building and mentorship 4) Familiarity with AI technology \\ \bottomrule
\end{tabular}}
\end{table*}